\documentclass[twocolumn,showpacs,aps,prl]{revtex4}

\usepackage{graphicx}%
\usepackage{dcolumn}
\usepackage{amsmath}
\usepackage{latexsym}

\begin {document}

\title
{
Weighted Scale-free Networks in Euclidean Space Using Local Selection Rule 
}
\author
{
G. Mukherjee$^{1,2}$ and S. S. Manna$^{1}$
}
\affiliation
{
$^1$Satyendra Nath Bose National Centre for Basic Sciences
    Block-JD, Sector-III, Salt Lake, Kolkata-700098, India \\
$^2$Bidhan Chandra College, Asansol 713304, Dt. Burdwan, West Bengal, India
}
\begin{abstract}

    A spatial scale-free network is introduced and studied whose motivation has been
    originated in the growing Internet as well as the Airport networks. We argue that
    in these real-world networks a new node necessarily selects one of its neighbouring
    local nodes for connection and is not controlled by the preferential attachment 
    as in the Barab\'asi-Albert (BA) model. This observation has been mimicked in our 
    model where the nodes pop-up at randomly located positions in the Euclidean space 
    and are connected to one end of the nearest link. In spite of this crucial difference 
    it is observed that the leading behaviour of our network is like the BA model.
    Defining link weight as an algebraic power of its Euclidean length, the
    weight distribution and the non-linear dependence of the nodal strength on 
    the degree are analytically calculated. It is claimed that a power law decay 
    of the link weights with time ensures such a non-linear behavior. Switching
    off the Euclidean space from the same model yields a much simpler
    definition of the Barab\'asi-Albert model where numerical effort grows linearly with
    $N$.

\end{abstract}

\maketitle

      Studying the structure of communication networks is important in their own right 
   because it helps in understanding the network, its traffic flow distribution which in turn helps
   making the communication process more efficient. Over last several years many real-world 
   networks are being studied with much interests. The examples vary from social 
   networks, electronic networks and biological networks \cite {Barabasi,Newman,Bose}. 
   An important subclass of these networks are spatial networks, i.e., those embedded in the 
   Euclidean space. Two most important examples of these networks are the electronic 
   communication network like the Internet \cite {Faloutsos,Pastor,Yook}, which is a transport 
   network of electronic data packets as well as the public transport system of 
   Airport networks \cite {Guimera,Barrat}.

      The common property of these two networks is their highly inhomogeneous structures. 
   This inhomogeneity is reflected in their nodal degree distribution (degree $k$ 
   of a node is the number of links attached to it) which is observed to follow the 
   power law distribution: $P(k) \sim k^{-\gamma_k}$. It was Barab\'asi and Albert (BA) who
   first recognized that indeed there are many other real-world examples in social and 
   biological networks having similar structures \cite {Barabasi1}. Since these networks
   lack a characteristic value for the nodal degree, they are called as the Scale-free 
   Networks (SFN) \cite {Barabasi,Barabasi1}.

      BA observed two key characteristics for these networks: (i) they are growing 
   networks (ii) existence of an inherent `rich get richer' mechanism which ensures that large
   degree nodes grow at higher rates. In BA model a network grows by addition
   of new nodes which get connected to the growing network using a linear attachment probability.
   In this paper we question the general necessity of this `linear attachment' rule. We 
   argue that at least for spatial networks like the Internet and the Airport Network existence 
   of such a rule for the expanding network seems to be highly implausible.

       If you buy a new computer and would like to connect it to the Internet what do 
   you do? If it is a home computer you use a modem and a dial-up telephone line to connect 
   it to the nearest router of the telephone company. If it is your office computer, it 
   gets a connection to the office router, which is eventually connected to the nearest 
   node of the Internet service provider (ISP) your organization had opted for. If your 
   office is a part of your university or organization, several routers are used and in 
   the autonomous systems level they are connected to the nearest ISP again. Therefore
   the new nodes of the Internet in any level are always connected to the existing local 
   nodes of the network. A person in Chile would hardly give any extra importance to large
   ISP hubs in Tokyo, Stuttgart or Chicago rather than small providers in his locality.
   Considering the whole world-wide Internet network it is apparent that the new nodes 
   pop up randomly in space and time without any spatial correlations.

      Similar arguments can be put forward for the airport network as well. A new airport
   in some remote city in some country is quite likely to be connected to the neighboring
   airport first by direct non-stop flights and very unlikely to have cross-country
   inter-continental flights to begin with. Recent studies on IATA airport network data
   also revealed that weights $w_{ij}$ of the links in this network either defined by the number of
   passenger seats or by the Euclidean distance between successive stops have non-trivial variations
   \cite {Barrat1}. The nodal strength defined as $s_i=\Sigma_j w_{ij}$ varies as $\langle 
   s(k) \rangle \propto k^{\beta}$ with $\beta > 1$. A number of models have been proposed 
   to put forward explanations for such non-linearity \cite {Wang,Bianconi,Goh}.

      In this paper we study a growing network on Euclidean space where new nodes are added
   one by one and are connected to the neighboring nodes of the growing network. We show 
   that even such a network has scale-free degree distribution. In addition, these networks
   have non-trivial dependence of average strength on the nodal degrees.

      A spatial network is grown on a two-dimensional space whose nodes are the points at 
   randomly selected positions within an unit square on the $x-y$ plane (with periodic 
   boundary conditions) \cite {Manna}. Let $\{(x_1,y_1),(x_2,y_2), ... , (x_N,y_N)\}$ 
   represent the co-ordinates of the of $N$ randomly distributed points within this space 
   where each co-ordinate is an i.i.d. number drawn within $\{0,1\}$. The growth of the network
   starts with a pair of nodes 1 and 2 connected by a single link. Nodes labeled 3 
   to $N$ are then introduced one by one and are connected to the growing network.

\begin{figure}[top]
\begin{center}
\includegraphics[width=6.5cm]{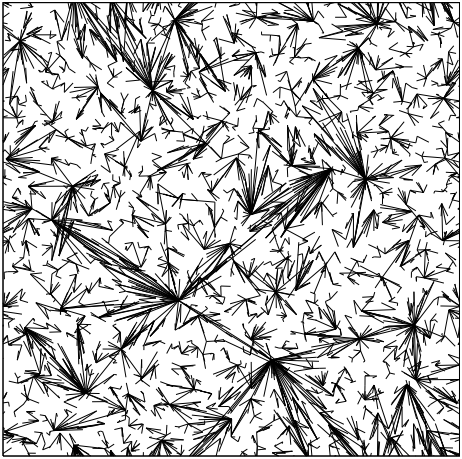}
\end{center}
\caption{
The picture of a network of $N=2^{12}$ nodes, each node is a randomly positioned
point on the unit square and is connected randomly to one end of the nearest link.
}
\end{figure}

      Let at some intermediate stage the network have $t$ links and $t+1$ nodes. For connecting 
   the $(t+2)$-th node to the network, the nearest link center is selected. One of the two end nodes 
   of the nearest link is then chosen with equal probability and is connected to the new node
   to create the $(t+1)$-th link (Fig. 1). This is executed by keeping the local informations
   into memory. The unit square area is divided into a lattice of size $\sqrt N \times \sqrt N$.
   As the network grows the serial numbers of the links whose centers are positioned within a 
   lattice cell are stored at the associated lattice site. To find out the nearest link center one
   starts from the cell of the $(t+2)$-th node and searches the lattice cells shell by shell 
   till the nearest link center is found out. When $t \sim N$ only the nearest shell needs to
   be searched.

      After the network has grown to $N$ nodes, the degree distribution $P(k,N)$ is calculated. From a 
   direct measurement of the slope of the $\log P(k,N)$ vs. $\log k$ plot the exponent $\gamma_k$ 
   is estimated to be $ 3.00 \pm 0.05$, which is close to BA value. Moreover, a scaling of 
   $P(k,N)$ is also studied (Fig. 2(a)):
\begin {equation}
P(k,N) \propto N^{-\eta} {\cal G}(k/N^{\zeta})
\end {equation}
   where $\eta = 3/2 $ and $\zeta = 1/2 $ are used to obtain the best data collapse giving $\gamma_k = 
   \eta / \zeta = 3$. The network so generated has a tree structure. However
   networks having more general structures with multiple loops can be generated 
   with $m$ links coming out from every incoming node and are attached in a similar way to the first
   $m$ neighboring links in increasing order. In Fig. 2(b) we show that degree distributions
   for $m=2$ and 3 can be scaled as $P(k,N)m^{-1.67} \propto N^{-\eta} {\cal G}(k/N^{\zeta})$.

\begin{figure}[top]
\begin{center}
\includegraphics[width=7.5cm]{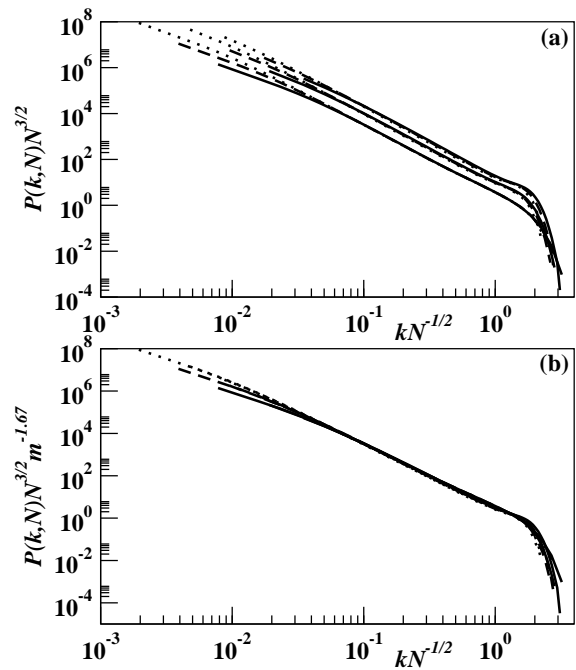}
\end{center}
\caption{
   (a) Finite-size scaling analysis of the degree distribution for three system
   sizes, $N = 2^{14}$ (solid line), $2^{16}$ (dashed line), $2^{18}$ (dotted line) and for three values of the 
   number of outgoing links $m$ = 1, 2 and 3 (from left to right). (b) Same plot as in (a) but
   scaled with $m^{-1.67}$.
}
\end{figure}

      The link lengths are also measured and their probability distribution is calculated
   as studied in \cite {Manna}. Let ${\cal D}(\ell)d\ell$ denote the probability that
   a randomly selected link has a length between $\ell$ and $\ell + d\ell$. For a given
   Poisson distribution of $N$ points let us first calculate the first neighbor distance distribution.
   Consider a point P at an arbitrary position. The probability that its first neighbor is situated at a distance
   within $r$ and $r+dr$ (which can be done in $N-1$ different ways) and all other $N-2$ 
   points are at distances greater than $r+dr$ is:
\begin {equation}
P(r)dr = (N-1) \hspace*{0.1 cm} 2\pi r dr \hspace*{0.1 cm} (1-\pi r^2)^{N-2}
\end {equation}
   In the limit of $N \to \infty$ it can be approximated that $N-1 \approx N-2 \approx N$ and since
   the average area per point decreases as $1/N$, $\pi r^2$ is very small compared to 1. Therefore
   $(1-\pi r^2)^{N-2}$ is approximated as $\exp (-\pi N r^2)$. Therefore in the limit of $N \to \infty$ the
   the probability density distribution is:
   $P(r) \approx 2\pi N r \hspace*{0.1 cm} \exp (-\pi N r^2)$
   or in the scaling form:
\begin {equation}
P(r)/\sqrt{N} \propto [r\sqrt {N}] \hspace*{0.1 cm} \exp (-\pi [r\sqrt {N}]^2)
\end {equation}
   where the scaling length $1/\sqrt {N}$ is the linear extent of the average area per node.

   In our case the number of nodes $N$ in the system is not fixed rather grows
   with time. Therefore the link lengths $\ell$ also decrease as time progresses. After some time
   the total collection of links has a mixture of many different
   lengths. Since intially there were only few nodes the early links are very large and may be
   as big as the box size, where 
   as the last few links are very small and have lengths of the order of $\ell_o \sim 1/\sqrt {N}$. 
   Effect of the mixing has been two fold as observed numerically. The distribution of
   small link lengths (later stage) up to $\ell_m \approx 2.5\ell_o$ is different from Eqn. 3 but still follows a 
   scaling form:
\begin {equation}
{\cal D}(\ell,N) \propto \sqrt{N} {\cal F}(\ell \sqrt {N}).
\end {equation}
   The scaling function fits very well to the generalized Gamma distribution:
   ${\cal G}(x) = Ax^{a}\exp(-bx^c)$
   where, $A \approx 12.6$, $a \approx 1.4$, $b \approx 3.11$ and $c \approx 0.83$ (Fig. 3(a)).
   On the other hand the long early links contribute to a power 
   law tail as ${\cal D}(\ell) \sim \ell^{-\gamma_w}$ for $\ell_m < \ell < \sqrt 2/2$ with $\gamma_w
   \approx 3.0$ (Fig. 3(b)). 

\begin{figure}[top]
\begin{center}
\includegraphics[width=7.5cm]{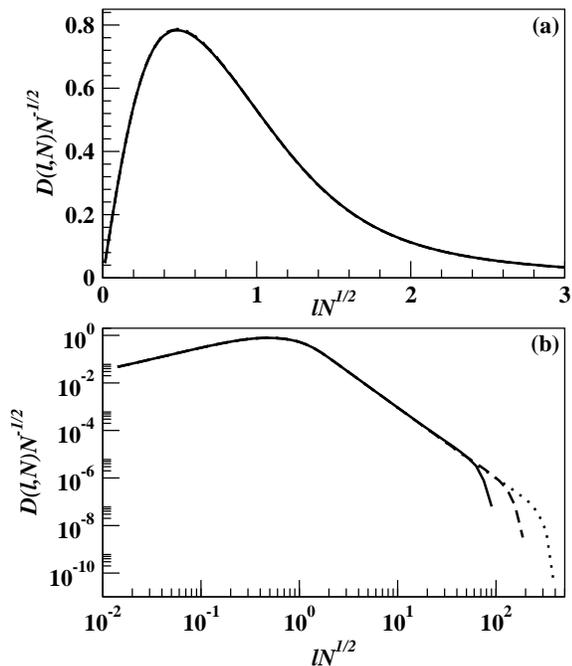}
\end{center}
\caption{
   Finite-size scaling analysis of the link length distribution
   $D(\ell,N)$ for three system sizes, $N = 2^{14}$ (solid line), $2^{16}$ (dashed line), $2^{18}$ (dotted line).
   Fig. (a) shows the plots on a linear scale where as in (b) the double logarithmic scale
   has been used for the same data but binned on exponential scale.
}
\end{figure}

      Like weighted airport network \cite {Barrat1} the strength $s_i$ of a node
   $i$ is measured as the sum of the Euclidean lengths of all links meeting at $i$:
   $s_i = \Sigma_j \ell_{ij}^{\alpha}$, $\ell_{ij}$ being the length of the link between 
   the nodes $i$ and $j$ and $\alpha$ is a continuously tunable parameter. 
   This parameter generalizes the model to take care of situations 
   where the link weight may even vary non-linearly with the Euclidean distance. E.g., 
   the route that an aircraft takes for flying between two airports
   is quite often greater than the length of the geodesic path between them.
   For example the ASIANA airlines flies from Seoul to New Delhi through the air space of Bangladesh.
   Also in Mobile ad hoc network (MANET) the power $P_i$ required to maintain the range 
   $R_i$ of transmission of each mobile element varies nonlinearly with the range: 
   $P_i \propto R_i^{\alpha}$, with $ 1 < \alpha < 6 $.

\begin{figure}[top]
\begin{center}
\includegraphics[width=7.5cm]{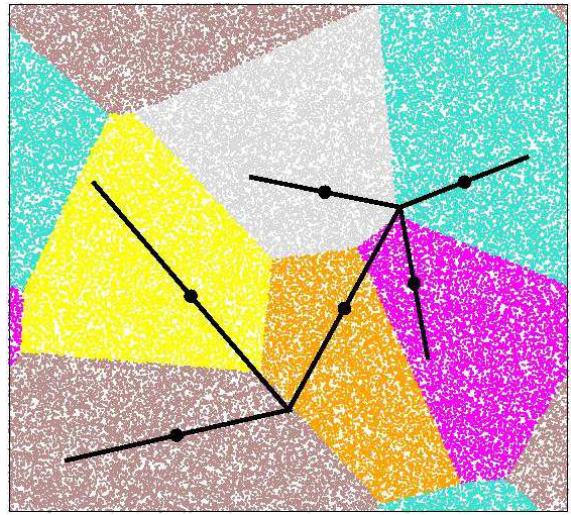}
\end{center}
\caption{
   (Color on-line) On an unit square a six link network is drawn and centers of the links
   are marked. The whole space is then partitioned into six different Voronoi cells around 
   these centers. Cells are marked by dropping $2^{16}$ randomly selected points in this 
   space and coloring all points in a cell by a particular color which are nearest to the 
   corresponding link center.
}
\end{figure}

   We first observe that given a connected network of $t$ links, the whole space is
   partitioned into $t$ non-overlapping Voronoi cells, each cell surrounds the center of
   a link \cite {Voronoi} (Fig. 4). The center of the link is at a minimum distance from 
   all points within this cell. The probability that a randomly selected point is within 
   a particular cell is equal to the area of the cell. Since different Voronoi cells have 
   different areas the probability of selecting a link center is in general non-uniform. For 
   a two-dimensional Poisson Voronoi tessellation the cell sizes follow a Gamma distribution 
   whose width scales as $1/t$ \cite {Szabo}. Therefore though for finite $t$ the cell 
   sizes are non-uniform, for $t \to \infty$ the cell size distribution 
   is similar to a Delta function implying that all cells have uniform size. In this
   limit the probability that a particular node of degree $k$ will be linked is
   $k/2$ times the cell size - which gives rise to the linear preferential attachment
   as in BA model. Therefore for a very large network ($N \to \infty$) in our model
   the node selection probability is different from that in BA model at early times ($t$ small)
   where as it asymptotically converges to the BA preferential rule in the limit of $t \to \infty$.
   We conclude that the leading behaviour of our spatial network model is like the BA model.
   We have already seen that even for finite $N$ the degree exponent $\gamma_k \approx 3.0$ 
   as in the BA model.

      Let us assume that the area of every Voronoi cell in the network with $t$ links is
   uniform and is equal to 
   $1/t$. This implies that the $i$-th node with degree $k_i(t)$ has probability $k_i(t)/(2t)$ to 
   be linked with the new $(t+2)$-th node. The factor 2 comes from the fact that one of the
   two end nodes of every link is selected with equal probability. Therefore 
   $dk_i(t)/dt = k_i(t)/(2t)$ and $k_i(t) 
   = (t/i)^{1/2}$, exactly similar to the BA model resulting $P(k) \propto k^{-3}$.

\begin{figure}[top]
\begin{center}
\includegraphics[width=7.5cm]{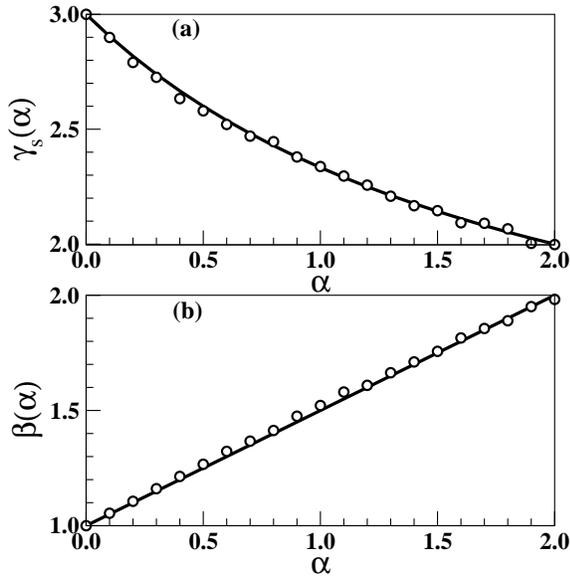}
\end{center}
\caption{Verification of the formulas for 
(a) $\gamma_s(\alpha) = 1+4/(2+\alpha)$ and (b) $\beta(\alpha) = 1+\alpha/2$ 
(solid line) with the numerical results (circles) for
$N=2^{14}$.
}
\end{figure}

      Again, since the area of each cell is $1/t$ the typical length $\ell_{ij}$ of the
   $t+1$-th link to be connected to the node $i$ is proportional to $(k_i(t)/t)^{1/2}$. 
   Therefore the rate of increase of the strength of the $i$-th node is:
\begin {equation}
ds_i(t)/dt = (dk_i(t)/dt).(\ell^{\alpha}_{ij}) \propto (k_i(t)/t)^{1+\alpha/2}
\end {equation}
   which is proportional to $(ti)^{-1/2-\alpha/4}$. On integration over $t$ from $t=i$ to $t$:
\begin {equation}
s_i(t)-s_i(i) \propto (\frac{t^{1/2-\alpha/4}}{i^{1/2+\alpha/4}}-i^{-\alpha/2}).
\end {equation}
   The value of $s_i(i)$ is estimated by the average strength of the $i$-th node when it 
   was introduced and connected to an arbitrary previous node $j, j=1,i-1$
\begin {equation}
   s_i(i) 
\propto i^{-1/2-\alpha/4}
\end {equation}
   When $t$ is large in Eqn. (4) the term $i^{-\alpha/2}$ is ignored
\begin {equation}
s_i(t) \propto {i^{-1/2-\alpha/4}}[t^{1/2-\alpha/4}+c]
\end {equation}
   Writing $i$ in terms of $s_i(t)$
\begin {equation}
i \propto [t^{1/2-\alpha/4}+c]^{4/(2+\alpha)}[s_i(t)]^{-4/(2+\alpha)}
\end {equation}

Since the serial number $i$ and the strength $s$ are both nodal quantities
they are dependent and their probability distributions obey $P(s,t)ds = - P(i,t)di$, the
negative sign is because as $i$ increases $s$ decreases. Using $P(i,t)=1/t$ we get
the time dependent probability distribution of the nodal strengths as:
\begin {eqnarray}
P(s,t) & = & -\frac{1}{t}\frac{di}{ds} \nonumber \\
       & \propto & [t^{1/2-\alpha/4}+c]^{4/(2+\alpha)}
             [\frac{4}{2+\alpha}s^{-1-4/(2+\alpha)}]\nonumber \\
       & \propto & t^{-\frac{2\alpha}{2+\alpha}}s^{-1-\frac{4}{2+\alpha}}
\end {eqnarray}
   Therefore $\gamma_s(\alpha) = 1+4/(2+\alpha)$ and using the general relation 
   $\gamma_s=\gamma_k/\beta+1-1/\beta$ \cite {Barrat} and using $\gamma_k=3$ we get
\begin{equation}
   \beta(\alpha)=1+\alpha/2.
\end{equation}
   The exponent of the distribution of $w_{ij}=\ell_{ij}^{\alpha}$ varies as $\gamma_w(\alpha)=1+2/\alpha$.
   A number of checks have been done to verify these results. For a large network with
   $N=2^{14}$ the strength distribution and its average are calculated for 21 different 
   values of $\alpha$ equally spaced between 0 and 2. Results are found to be quite 
   consistent with the above formulas of $\gamma_s(\alpha)$ and $\beta(\alpha)$ (Fig. 5). 

\begin{figure}[top]
\begin{center}
\includegraphics[width=7.5cm]{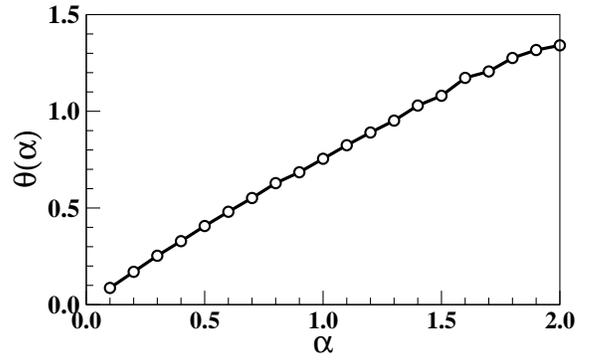}
\end{center}
\caption{Variation of the exponent $\theta(\alpha)$ with $\alpha$ for $N=2^{14}$.
}
\end{figure}

   Therefore the message is that a nontrivial value of $\beta > 1$ 
   indeed can be obtained in a weighted network when the link weights 
   decrease systematically as time elapses. In the above model the link weight
   decreases as $(it)^{-\alpha/4}$. This 
   is consistent with the airport network data where early airports in big cities like London, 
   Paris etc. have very high strengths. Because of the fact that these early airports
   still survive over more than a century implies that they are connected with strong links of 
   high passenger traffics as well as long distance links. Therefore if the early links 
   have maximal weights and if they decrease inversely with a power of time, that can result 
   $\beta > 1$. 

   A simple way to check this idea is to study the generalized BA model itself, starting from
   a single link. The weights of the links are assigned by hand: A link which becomes a part
   of the network at time $t$ carries a weight $t^{-\alpha}$. Then if the degree of the largest
   hub $\langle k_{max} \rangle$ grows as $t^{\mu}$ ($\mu=1/2$ for the BA model) then
   it can be shown that:
   for $\alpha < \mu$, $\beta(\alpha) = 1$ and
   for $\alpha > \mu$, $\beta(\alpha) = \alpha /\mu$. 
   The weight distribution exponent varies with $\alpha$ as $\gamma_w(\alpha)=1+1/\alpha$.
   We observe that these results are very much consistent with the results in \cite {Bianconi}.
   In this model each link comes in with a constant weight and at the same time $m'$ randomly 
   selected links gain additional weights $w_o$ at each time step. As a result old links 
   increase their weights as time increases. On simulating the Bianconi model network of size 
   $N=2^{14}$ we calculated that the average weight of a link introduced at time $i$ decreases as 
   $i^{-\alpha}$ which is exactly their result that the weight of a link increases with time 
   as $(t/i)^{\alpha}$. In comparison to the conclusion of \cite {Goh} that a global reorganization of
   weights (as in \cite {Goh} and \cite {Bianconi}) yields $\beta > 1$, we claim that ultimately
   the global reorganization or any other dynamics has to ensure a power law decay of the link
   weights like $t^{-\alpha}$ to achieve $\beta > 1$. Similar conclusion has been drawn in 
   \cite {Bianconi}.

   To study the correlation between the nodal degrees and the link weights Barrat et. al. \cite {Barrat} 
   calculated how the product of the degrees $k_ik_j$ of the two end points of a link varies 
   with the link weight $w_{ij}$
\begin {equation}
\langle w_{ij} \rangle \propto (k_ik_j)^{\theta(\alpha)}
\end {equation}
   A direct calculation of $\langle w_{ij} \rangle$ averaged over different configurations 
   generated for different values of $\alpha$ for our spatial network shows that variation is 
   almost linear (Fig. 6).

      We further observe that our model can be generalized when the new node selects 
   one of the $t$ links randomly with a probability $d_n^{\delta}/\Sigma_{n=1}^td_n^{\delta}$
   where $\delta$ is a continuously tunable parameter. Therefore in the limit of $\delta \to 
   -\infty$ we retrieve the above model. On the other hand for $\delta=0$ the underlying Euclidean 
   space is switched off and the networks are generated in a much simpler way: To add the $(t+1)$-th 
   link, a link $j$ is selected out of $t$ links with uniform probability $1/t$. The new $(t+2)$-th 
   node is then randomly connected to one end of $j$ with probability 1/2. Therefore a node of 
   degree $k$ has the probability $k/(2t)$ to be linked with the new node which clearly satisfies 
   the requirement of linear attachment probability of the BA model. Therefore we claim that the 
   algorithm: {\it the new node selects one of the links with uniform probability and gets 
   connected to its one end with probability 1/2} is exactly the BA algorithm. Computationally 
   there is a lot of advantage, it takes a CPU linearly proportional to $N$ as already observed 
   in \cite {Newman,Krapivsky,Dorogov}. The network so generated clearly has a tree structure. However, 
   loops can also be generated easily by connecting each new node with $m$ distinct links, each 
   link is attached with the same rule. The resulting network is exactly the BA network with $m$ 
   outgoing links from each node.

     Finally it is observed that the case of the non-linear attachment rule in BA model when the 
   probability of attachment varies as $k^{\epsilon}$ cannot be obtained by generalizing this 
   model. Let the two end nodes of the randomly selected link have degrees $k_1$ and $k_2$ 
   then depending on their degrees one of them is selected with the following probabilities:
   \begin {equation}
   p(k_1) = k_1^{\epsilon}/(k_1^{\epsilon}+ k_2^{\epsilon})
   \quad {\rm and} \quad
   p(k_2) = k_2^{\epsilon}/(k_1^{\epsilon}+ k_2^{\epsilon})
   \end {equation}
   with $p(k_1) + p(k_2)$ = 1. The BA model corresponds to $\epsilon=0$. However on reducing 
   $\epsilon$ gradually through its negative values, the node with the smaller degree gets 
   more preference to be connected. Therefore in the limit of $\epsilon \to -\infty$ the node
   with smaller degree is always selected with probability one. Even in this limit the network
   has a branched structure because if the two end nodes have equal degrees, any one of them 
   gets connected with the new node with probability 1/2. On the other hand when $\epsilon >
   0$, the node with a larger degree gets more weight and in the limit of $\epsilon \to 
   \infty$ the end node with larger degree always gets the new connection resulting a star
   like structure.  When $\epsilon $ decreases below zero, the degree distribution becomes 
   stretched exponential like: $P(k, \epsilon <0) \sim \exp(-ak^{b(\epsilon )})$ where the 
   exponent $b(\epsilon )$ has a continuous variation with $\epsilon $, expected to reach 1 
   as $\epsilon \to -\infty$ and 0 when $\epsilon \to 0$. Our numerical estimates for 
   $b(\epsilon )$ for different values of $\epsilon $ fit very well to the form $b(\epsilon )
   = a_o(-\epsilon )^{\nu}-b_o$ where the constants are estimated to be $a_o \approx 1.19$, 
   $\nu \approx 0.14$ and $b_o \approx 0.85$.

      To summarize, we argued that in the growing Internet as well as in the Airport network 
   it is more likely that the new nodes get their connections in the local neighborhood.
   Indeed a spatial scale-free network is grown using the criterion of local selection
   rule. This network shows a non-linear dependence of the nodal strength on the degree.
   We conjectured that irrespective of whatever the dynamics may be, the non-linearity is
   the result of a power law decay of the link weight with time. When the same network is 
   constructed without the underlying Euclidean space it gives a very efficient algorithm to 
   generate the Barab\'asi-Albert network.

   GM thankfully acknowledged facilities at S. N. Bose National Centre 
   for Basic Sciences. 

\leftline {Electronic Address: manna@bose.res.in}


\begin{thebibliography}{90}

\bibitem {Barabasi} R. Albert and A.-L. Barab\'asi, Rev. Mod. Phys. {\bf 74}, 47 (2002).

\bibitem{Newman}  M. E. J. Newman, SIAM Review {\bf 45}, 167 (2003).

\bibitem {Bose} I. Bose in {\it Frontiers in Biophysics} p87 ed. by T. P. Singh
and C. K. Dasgupta , Allied Publishers, New Delhi (2005).

\bibitem {Faloutsos} M. Faloutsos, P. Faloutsos and C. Faloutsos, Proc.
               ACM SIGCOMM, Comput. Commun. Rev., {\bf 29}, 251 (1999).

\bibitem {Pastor} R. Pastor-Satorras, A. Vazquez and A. Vespignani, Phys. Rev. Lett. {\bf 87}, 258701 (2001).

\bibitem {Yook} S. H. Yook, H. Jeong and A.-L. Barab\'asi, Proc. Natl. Acad. Sci. (USA) {\bf 99}, 13382 (2002).

\bibitem {Guimera} R. Guimera and L. A. N. Amaral, Eur. Phys. Jour. B, {\bf 38}, 381 (2004).

\bibitem {Barrat} A. Barrat, M. Barth\'el\'emy, R. Pastor-Satorras, A. Vespignani, Proc. Natl. Acad. Sci. (USA),
{\bf 101}, 3747 (2004). 

\bibitem {Barabasi1} A.-L. Barab\'asi and R. Albert, Science, {\bf 286}, 509 (1999).

\bibitem {Barrat1} A. Barrat, M. Barth\'el\'emy and A. Vespignani, J. Stat. Mech. (2005) P05003.

\bibitem{Wang} W. X. Wang, B. H. Wang, B. Hu, G. Yan and Q. Ou, Phys. Rev. Lett., {\bf 94}, 188702 (2005).

\bibitem{Bianconi} G. Bianconi, Eurphys. Lett. {\bf 71} 1029 (2005).

\bibitem{Goh} K.-I. Goh, B. Kahng and D. Kim, Phys. Rev. E {\bf 72}, 017103 (2005).

\bibitem {Manna} S. S. Manna and P. Sen, Phys. Rev. E {\bf 66}, 066114 (2002).

\bibitem {Voronoi} G. Voronoi, J. Reine Angew. Math. 134, 198 (1908).

\bibitem {Szabo} F. J\'arai-Szab\'o and Z. N\'eda, arXiv:cond-mat/0406116.

\bibitem{Krapivsky} P. L. Krapivsky and S. Redner, Phys. Rev. E. {\bf 63}, 066123 (2001).

\bibitem {Dorogov} S. N. Dorogovtsev and J. F. F. Mendes, {\it Evolution of Networks},
Oxford University Press, 2003.

\end{thebibliography}
\end {document}